\documentclass{aa}
\input epsf

\newcommand\apj {ApJ}
\newcommand\aj {AJ}
\newcommand\apjs {ApJS}
\newcommand\mnras {MNRAS}
\newcommand\aap {A{\&}A}
\newcommand\aaps {A{\&}ASS}
\newcommand\code {SPECTRAL}

\newcommand{\beq}{\begin{equation}}
\newcommand{\eeq}{\end{equation}}

\newcommand{\gax} {\ifmmode{_>\atop^{\sim}}\else{${_>\atop^{\sim}}$}\fi}
\newcommand{\lax} {\ifmmode{_<\atop^{\sim}}\else{${_<\atop^{\sim}}$}\fi}

\begin{document}

\title{\code: New Evolutionary Synthesis code. An application to the 
irregular galaxy NGC 1560}

\author{Gerardo A. V\'azquez\inst{1}, Leticia Carigi\inst{1}, 
        \& J. J. Gonz\'alez\inst{1}}

\offprints{Gerardo A. V\'azquez}

\institute {Instituto de Astronom\'\i a, Universidad Nacional Aut\'onoma de 
M\'exico,
            A.P. 70-264, D.F. 04510, M\'exico \\
            email: gerar, carigi, jesus@astroscu.unam.mx}

\authorrunning{G.A. V\'azquez, L. Carigi \& J.J. Gonz\'alez}
\titlerunning{{\code}: Spectro-Chemical Evolution for NGC 1560}

\date{Received date; accepted date}

\abstract{
We have developed a new evolutionary synthesis code, which incorporates  
the output from chemical evolution models. We compare results of this new 
code with other published codes, and we apply it 
to the irregular galaxy NGC 1560 using sophisticated chemical 
evolution models. The code makes important contributions in two areas: 
a) the building of synthetic populations with time-dependent star formation 
rates and stellar populations of different metallicities; b) the extension of 
the set of stellar tracks from the Geneva group by adding the AGB phases for 
$m_i/M_\odot \geq 0.8$ as well as the very low mass stars. Our code predicts 
spectra, broad band colors, and Lick indices by using a 
spectra library, which cover a more complete grid of stellar parameters. The 
application of the code with the chemical models to the galaxy NGC 1560 
constrain the star formation age for its stellar population around 10.0 Gy.
}
\maketitle

\section{Introduction}

\label{sec:int}

The study of the history of the evolution of galaxies includes three important 
issues: {\it spectral}, {\it dynamical}, and {\it chemical 
evolution}. Since these three different parts of the galaxy evolution are very 
difficult to study in simple models. Different authors has been focused to 
one of these parts. Although several efforts have been devoted by different 
authors (for example \cite{tinsley76} and \cite{vazdekis96}) to build 
complete spectro-chemical evolution codes. It has been difficult to 
incorporate the new releases in stellar models (structure evolution, stellar 
yields and stellar atmospheres) to the complexity of galaxy evolution models. 
The last reason has resulted in the division of 
the study of the spectro-chemical galaxy evolution in two 
complementary ways. While the chemical evolution models are the independent 
variable in the complete spectro-chemical evolution, the spectral evolution 
codes, which use population synthesis or evolutionary synthesis are left to 
use simple star formation histories to constrain the results and to be 
independent models. 

In this way, we find sophisticated chemical evolution models. But, the most 
part of spectral evolution models consider the star formation in a simple 
star-burst, and a complicated star formation history can be modeled by many 
star-bursts of different metallicities and star formation rates. Thus, instead 
of studying a broad range of galaxies using simple star-burst 
models, it is preferable to develop spectro-chemical evolution models of 
neighboring galaxies using constrains 
to model the systems, and after that use the models to galaxies 
at higher red-shifts.

The new code ({\code}) presented here has two important contributions to the 
evolutionary synthesis models: the first one is the consideration of a more 
complete set of evolutionary tracks from the Geneva group, adding tracks from 
\cite{chabrier97} to the very low mass range; the second one is the 
construction of synthetic populations following the chemical evolution model 
in order to derive the spectral properties for systems with arbitrary 
enrichment history.

In this paper we initially use simple star formation scenarios to test the new 
evolutionary synthesis code, and then we proceed to more complete chemical 
evolution models to constrain the age of stellar population in the irregular 
galaxy NGC 1560.

In Section \ref{sec:trac} we illustrate how the tracks are assembled to 
built a complete set of tracks for either the high mass loss rates, or 
the normal mass loss rates, and we describe the spectra library used in our 
code. In Section \ref{sec:syn}, we describe the process followed to built the 
evolutionary synthesis code, and the transformation to the observational plane 
of the variables. In Section \ref{sec:models} we test and compare 
{\code}  with others codes commonly used in the field and we present a first 
application by predicting spectral variables for the irregular galaxy NGC 
1560. Finally, we discuss our conclusions.

\section{Stellar Tracks}

\label{sec:trac}

We have incorporated all the tracks from the Geneva group: \cite{schallI}, 
\cite{schar93aII}, \cite{charb93III}, \cite{schar93bIV}, \cite{meynet94V}, 
\cite{charb96VI}, \cite{mowlav98VII} and \cite{charb99VIII}; together with 
tracks from Chabrier \& Baraffe (1997, CB) for the very low mass stars, in 
order to assemble a complete dual grid of evolutionary tracks for the spectral 
code. One set with higher stellar mass loss and the other with normal mass 
loss as showed in Table \ref{tbl-1}. {\bf All these tracks are published in 
such way that each evolutionary point over each track has an equivalent 
evolutionary point in the rest of the tracks} except for tracks from CB. This 
can be seen from the first line in the Table \ref{tbl-1} because the tracks 
have the same number of evolutionary points (51). This is a very important 
aspect, because it is the basis of our evolutionary synthesis code.

\begin{table}
\caption{Ingredients of the assemble of tracks.} \label{tbl-1}
\begin{center}\scriptsize
\begin{tabular}{lcccccc}
Tracks  & Phases & Points & Low & High & Metalli- \\
Group & & & Mass & Mass & cities \\
\hline
\hline
I, II & MS, RGB, & 51 & 0.8 & 120.0 & $a$ \\
III, IV & HB${^1}$, AGB${^1}$  & & & & \\
V & MS, RGB, & 51 & 12.0 & 120.0 & $b$ \\
  & HB, AGB & & & &  \\
VI & MS, RGB, & 54 & 0.8 & 1.7 & $c$ \\
  & HB, AGB & & & &  \\
VII & MS, RGB, & 51 & 0.8 & 60.0 & $d$ \\
  & HB${^1}$, AGB${^1}$ & & & &  \\
VIII & MS, RGB, & 51 & 0.4 & 1.0 & $e$ \\
CB & PMS, MS, & 4 & 0.075 & 0.8 & $f$ \\
\hline
\hline
\end{tabular}

$a$ $Z= 0.001$, 0.004, 0.008, 0.02, 0.04 \& $\dot M$ $\propto Z^{0.5}$

$b$ $Z= 0.001$, 0.004, 0.008, 0.02, 0.04 \& $\dot M$ $\propto 5.0 \times Z^{0.5}$

$c$ $Z= 0.001$, 0.02 \& $\dot M$  $\propto Z^{0.5}$

$d$ $Z= 0.1$ \& $\dot M$  $\propto Z^{0.5}$

$e$ $Z= 0.001$, 0.02 \& $\dot M$  $\propto Z^{0.5}$

$f$ $Z= 0.0002$, 0.00063, 0.001, 0.002, 0.0063, 0.02 \& $\dot M = 0$

\footnote{1}{Only for stars $m_i/M_\odot > 1.7$}
\end{center}
\end{table}

We have chosen tracks with overshooting to be assembled by using linear and 
bilinear logarithmic interpolation when it is possible and the final result 
is shown in Table \ref{tbl-2}. The process to assemble the tracks is 
described below, and all parameters are logarithmic. 

\begin{table}
\caption{Main features from the new set of tracks.} 
\label{tbl-2}
\begin{center}\scriptsize
\begin{tabular}{lcccccc}
Tracks  & Phases & Points & Low & High & Metalli \\
Group & & & Mass & Mass & cities \\
\hline
\hline
 Geneva & PMS- & 84 & 0.08 & 120.0 & $a$ \\
 + Lyon & AGB${^1}$ & & & & \\
 Geneva & PMS- & 84 & 0.08 & 120.0 & $b$ \\
 + Lyon & AGB${^1}$ & & & & \\
\hline
\hline
\end{tabular}

$a$ $Z= 0.001$, 0.004, 0.008, 0.02, 0.04, 0.1 \& $\dot M$ $\propto Z^{0.5}$

$b$ $Z= 0.001$, 0.004, 0.008, 0.02, 0.04 \& $\dot M$ $\propto 5.0 \times Z^{0.5}$

$^1$PMS-AGB means PMS, MS, RGB$^2$, HB$^2$, AGB$^2$

\footnote{1}{Only for stars $m_i/M_\odot > 0.7$}
\end{center}
\end{table}

First of all, we completed phases covered at each metallicity. 
This means that tracks, which do not have post-main sequence 
evolution for masses in the range $m_i/M_\odot \leq 1.7$ ($m_i$ is the 
initial mass, mass on main sequence) were completed to get 
51 evolutionary points covering phases from the post-main sequence RGB, HB, 
and AGB.

We have used a IDL code to visualize each interpolation 
in tracks, using weights for linear interpolations defined:

\begin{equation}
\quad{wf_1 = {(f_x-f_{j-1})\over (f_j-f_{j-1})}, \ \ \ wf_0 = 1.0 - wf_1}, 
\label{wmass}
\end{equation}
\noindent where $f_x$ is the desired variable for which the properties are 
calculated when we interpolate in mass, metallicity or evolutionary phase. The 
sub-indices mean adjacent variables in tracks for the interpolation. New 
variables ($V$) were calculated using the equation:

\begin{equation}
\quad{V(f) = V(f_{j-1})\times{wf_0} + V(f_j)\times{wf_1}}.
\end{equation}

Phases beyond RGB were not present for stars in the mass range 
$1.0 \leq m_i/M_\odot \leq 1.7$, and  post-main sequence phases were not 
present for stars in $0.8 \leq m_i/M_\odot < 1.0$ at some metallicities. The 
first step in the new set of tracks was to complete tracks up to RGB for those 
masses in $0.8 \leq m_i/M_\odot \leq 1.7$ using the last equations. 

The second step was to add phases as HB and AGB for the range in masses 
$0.8 \leq m_i/M_\odot \leq 1.7$. To do the interpolation and extrapolation 
in this phases we used tracks from two sets at metallicities $Z =$ 0.001 and 
0.02 mentioned in the tracks group VI in Table \ref{tbl-1}. In this way, it 
was necessary to interpolate and extrapolate this phases in metallicity. We 
followed in this part the shape of tracks in groups I, II, III, IV, and VII 
for the RGB to add the other two phases HB and AGB in the next form:

We completed the evolutionary points for the RGB in the tracks group VI up to 
30 instead of original 3 using the fact that luminosity grows in this phase 
linearly as explained in I. We did the same for the rest of 
tracks for masses $m_i/M_\odot \geq 1.7$ in groups I, II, III, IV, and VII to 
get 81 evolutionary points for tracks instead of 51 in the first group. 

Furthermore, to interpolate and extrapolate the post-RGB phases for 
the rest groups in metallicity we use the next method:

The metallicity against the shape of each stellar variable ($age$, 
{\it current mass} $(m_c)$, $L/L_\odot$, and $T_{eff}$) is fitted for linear 
functions $f_1(Z)$ and $f_2(Z)$ for both groups of tracks I, II, III, IV, VII, 
and VI, respectively, by using the metallicities at which both groups 
coincide, in this case $Z = $0.001 and 0.02. 

Then, we defined $Q(Z)=f_1(Z)/f_2(Z)$, as the factor to convert the shape of 
$f_1(Z)$ in to $f_2(Z)$ shape, or vice versa. Actually, we used 
the last point in RGB for the first tracks group as the first point in the new 
set of tracks and we interpolate in metallicity $\Delta(age)$, 
$\Delta(m_c)$, $\Delta(L/L_\odot)$, $\Delta(T_{eff})$ from the second group of 
tracks with the last method. Then, the second point in the new set of tracks 
is the last point in the RGB plus the first $\Delta$ interpolated, and so on.

The final step to assemble the tracks was to incorporate the very low mass 
stars by using models from CB. An interpolation in metallicity has been done 
to put these tracks as the same metallicities than Geneva tracks. Then, we 
have identified over the four evolutionary points of tracks, ZAMS points and 
pre-main sequence points to obtain in the full new set of tracks 84 
equivalent evolutionary points from pre-main sequence up to AGB.

\begin{figure}
\vspace{0cm}
\epsfxsize=8.5cm 
\epsfbox{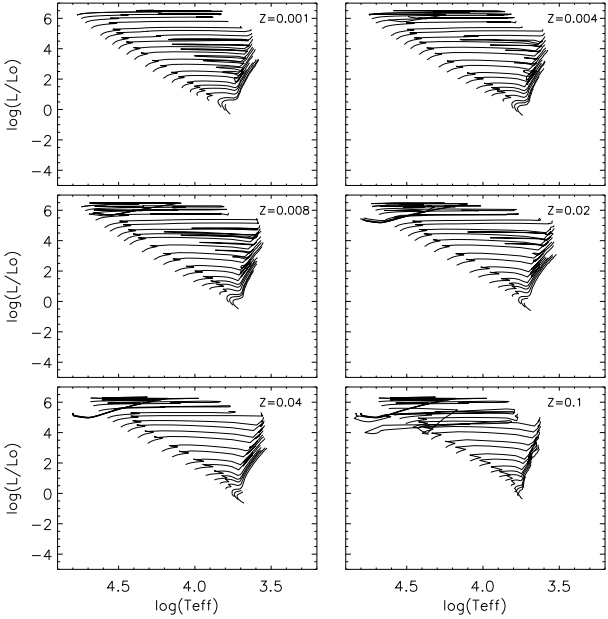} 
\vspace{0cm}
		 {\caption[]{Log($T_{eff}$) vs. log($L/L_\odot$). Tracks I, 
II, III, IV, and VII from Geneva group, for masses 0.8 - 120.0 M$_\odot$.}
\label{f1}}
\end{figure}

\begin{figure}
\vspace{0cm}
\epsfxsize=8.5cm 
\epsfbox{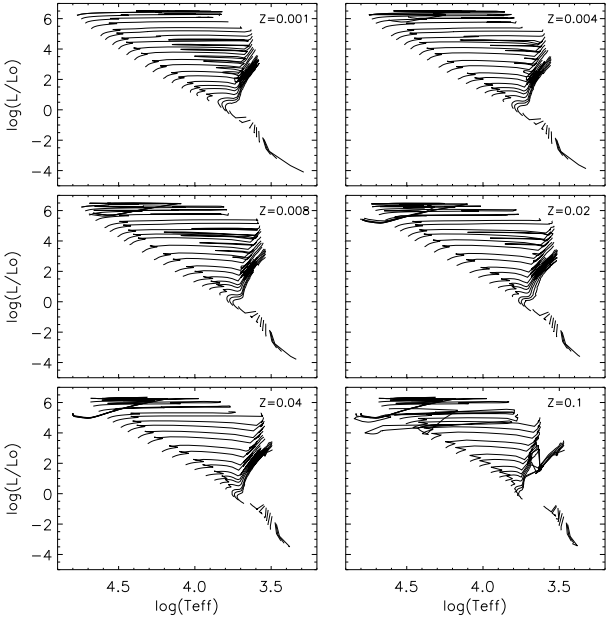} 
\vspace{0cm}
		 {\caption[]{Log($T_{eff}$) vs log($L/L_\odot$). New sets, 
of tracks, for masses 0.08 - 120.0 M$_\odot$.}
\label{f2}}
\end{figure}

For comparison we show in the Figure \ref{f1} and Figure \ref{f2} Geneva 
tracks and the new set of tracks with new phases, and new stellar tracks for 
very low masses with normal mass loss rate, respectively.

We can see in Figure \ref{f2} the pre-main sequence phases included for masses 
$m_i/M_\odot \leq 0.7$. In the same way, tracks for $Z=0.1$ are very poorly 
interpolated because these tracks are originally incomplete, and have been 
completed with tracks with $Z=0.04$ for masses $m_i/M_\odot > 60$. Finally, 
it has been obtained two homogeneous grids to interpolate the stellar 
variables in time, mass, and metallicity either, one with normal mass loss 
($\dot M \propto Z^{0.5}$) and second one using tracks from the V group for 
masses in the range $12.0 \leq m_i/M_\odot \leq 120.0$ in the 
Table \ref{tbl-1} with high mass loss ($\dot M \propto 5 \times Z^{0.5}$), 
and stars in the range of $m_i/M_\odot < 12.0$ completed in the same way as 
the set with normal mass loss showed in Figure \ref{f2}.

\section{The Evolutionary Synthesis Code {\code}}
\label{sec:syn}

\subsection{The Input and Synthetic Stellar Populations}

The evolutionary synthesis code has been developed in such away that it is 
possible to use the output of chemical evolution models taking from these 
models the following parameters:

-Duration of star formation, $t_{SFH}$.

-Star Formation Rate, $\Psi(t)$.

-Metallicity of interstellar gas at time $t$, $Z(t)$.

The code needs the number of stars $n(m_i,t)$ formed at any time in the interval 
$m_l \leq m_i/M_\odot \leq m_h$ defined as follows:

\begin{equation}
n(m_i,t) = 
\mbox{$\Phi(m_i)\Psi(t)$}\\
\end{equation} 
where $m_l$, $m_h$ are the limits in mass and $\Phi(m_i)$ is the initial mass 
function. 

$Z(t)$ is the metallicity of gas at which a new stellar population is 
born. Therefore, the new stars have a different metallicity or at least 
equal, predicted by the chemical evolution model.

When we see the synthetic stellar population formed over the whole continuous 
star formation history from $t=0$ up to $t_{SFH}$ at $Age$ of 10.0 Gy for 
instance, we need to calculate how many stars are still alive since they have 
formed in each step of time with the SFR (in this case those stars in the tip 
of AGB). If $t_{SFH} = Age$ then the result of this procedure is that most of 
the stars formed in the first generation have died after 10.0 Gy, and just 
the very low mass stars are alive. The last generation formed at 10.0 Gy is 
complete and the age for this population ($A_s$) is zero. This means:

\begin{equation}
\quad{A_s = Age - t}
\end{equation}
where $t$ runs from 0.0 up to $t_{SFH}$ which could be the same as $Age$.

The final result at 10.0 Gy is a mix of populations after the chemical 
evolution  has regulated the SF.

One of the problems in the evolutionary synthesis codes is the oscillations 
in the evolution of spectral parameters, because the post main sequence phases 
are not well sampled, due to the faster evolution of stars after the hydrogen 
burning phase. As shown in the Section \ref{sec:trac}, the problem is solved 
by interpolating in phases (evolutionary points) over the post hydrogen 
burning evolution, and interpolating in time and mass over main sequence 
phases.

To calculate which stars are alive at $Age$ in the post-main sequence region, 
we first calculate the masses of stars at each evolutionary points since the 
turn off point up to AGB using relations $ta(Z,m_i,j)$ vs $m_i$ (where $ta$ 
is the age of each evolutionary point $j$ over each track and initial mass 
$m_i$). If those ages satisfy the condition

\begin{equation}
\quad{ta(Z,m_i,j) \leq A_s \leq ta(Z,m_i,j+1)},
\end{equation} 
this means, 70 ages every time step, which cover exactly important 
post-main sequence evolutionary phases. The interpolation in phases is 
extended to metallicity using equation (1) and (2) to obtain:

\begin{equation}
\begin{array}{l}
{ta(Z,m_i,j) = ta(Z_0,m_i,j)\times{wZ_0} +}\\
{\ \ \ \ \ \ \ \ \ \ \ \ \ \ \ \ \ \ \ ta(Z_1,m_i,j)\times{wZ_1}},\\
\end{array}
\end{equation}
which represents the age of each 70 evolutionary points $ta$, at required 
metallicity for all masses in tracks $m_i$. Masses at each 70 
evolutionary points at required $A_s$ are calculated using equation (1) 
for the time $ta(Z,m_i)$ in the following way:

\begin{equation}
\begin{array}{l}
{m_i(Z,ta) = m_i(Z,m_{i_0},ta)\times{wt_0} +}\\ 
{\ \ \ \ \ \ \ \ \ \ \ \ \ \ \ \ \ m_i(Z,m_{i_1},ta)\times{wt_1}}\\
\end{array}
\end{equation}
where weights in time ($wt_0$ y $wt_1$) use $ta(Z,m_i,j)$ and $A_S$.

To sample those masses in main sequence and pre-main sequence phases, we 
divide the mass range $m_l \leq m_i \leq m_{Toff}$, where $m_{Toff}$ is 
the mass in the turn off point, in 80 masses to get a 150 masses for the 
synthetic population for each time step of evolution.

Once obtained this masses, we can calculate the stellar properties like 
$m_c$, $L/L_\odot$, $T_{eff}$ interpolating in time, mass, and 
metallicity by using the equations (1) and (2) shown in Section 
\ref{sec:trac}.

\subsection{Transformation from theoretical to observational plane}

The stellar parameters in the synthetic population are transformed 
to the observable parameters by using the calibration of \cite{lejeuneI}, and 
\cite{lejeuneII} which includes low resolution spectral energy distribution's 
(SED's) and broad band colors. Table \ref{tbl-3} shows the grid of parameters 
covered by this library.

\begin{table}
\caption{Features of SED's and broad band colors calibration used in this 
work.} \label{tbl-3}
\begin{center}\scriptsize
\begin{tabular}{rc}
\hline
Para- & Lejeune et al. \\
meters & (1997, 1998) \\
\hline
\hline
$T_{eff}$ (K) & $(2000, 50000)$ K \\
log$g$ & $(-1.02, 5.5)$ \\
$[M/H]$ & $(-5.0, 1.0)$ \\
$\lambda\lambda$ (nm) & $(9.1, 160000)$ \\
n($\Delta\lambda$) & 1221 \\
Colors & 14 \\
\hline
\end{tabular}
\end{center}
\end{table}

Following the same procedure as in tracks, we interpolate linearly over this 
grid using the same equations described in the Section \ref{sec:trac}, but 
now for $[M/H]$, $T_{eff}$, and $log(g)$ (gravity), where $[M/H] = 
log(Z/Z_\odot)$, with $Z_\odot=0.02$. Besides the spectrum, and broad band 
colors, we have included the Lick indices \cite{wfgb94} using the routine 
developed by J.J. Gonzalez in \cite{chucho}.

To integrate the observable variables for all alive stars we have to 
transforms in light or flux those variables given in magnitudes like colors 
and indices. In this way, we use the formalism described in \cite{chucho} to 
integrate the Lick indices, and 

\begin{equation}
\quad{L_{f_1} = L_{f_0} \times 10^{-0.4 \times (f_1-f_0)}},
\end{equation}
where $f_1$ can be the $V$ filter, and $f_0$ the bolometric correction, 
for instance, and we follow the same procedure for the rest of filters using 
the rest of colors.

The spectral variables are weighted by the star number and the average values 
are:

\begin{equation}
\quad{<L_f> = {{\int_{0}^{Age}\int_{m_l}^{m(t)} L_f(Z,m',t')\,w(m',t')\,
dm'\, dt'} \over{\int_{0}^{Age}\int_{m_l}^{m(t)} w(m',t')\,
dm'\, dt'}}} 
\end{equation}
where the weights to obtain the average are defined 

\begin{equation}
w(m',t') = \left\{
\begin{array}{l}
\mbox{$n(m',t') \times F_C$\ \ \ \ \ \ \ For indices}\\
\mbox{$n(m',t') \times L_{bol}$\ \ \ \ \ \ For colors}\\
\mbox{$n(m',t')$\ \ \ \ \ \ \ \ \ \ \ \ \ \ \ For spectra,}\\
\end{array}
\right.
\end{equation}
and, $L_f(Z,m',t')$ is the stellar variable to integrate, $L_{bol}$ is the 
bolometric luminosity and $m(t)$ is the mass of star, which is alive at time 
$t$.

Finally, integrated variables are obtained by using :

\begin{equation}
\quad{<(f_1 - f_0)> = -2.5 \times log({<L_{f_1}> \over <L_{f_0}>})},
\end{equation}
for indices and colors, and 

\begin{equation}
\quad{<mag_f> = -2.5 \times log(<L_f>)},
\end{equation}
in the case of magnitude indices.

\section{Evolution Models}

\label{sec:models}

\subsection{Tests for the Evolutionary Synthesis Code}

We have used instantaneous star-burst models to test and compare our code 
with others commonly used. 

In the first part, we test colors and spectral indices; in the second part 
spectra for young and old populations.

\subsubsection{Colors and Spectral Indices}

For the first part we assume the next set of conditions:

An $\Phi(m_i) \propto m^{\alpha}$, with $\alpha=-2.35$, low mass limit, 
$m_l=0.1$, and high mass limit, $m_h=100.0$; and for {\code}, 
we use Geneva tracks with normal mass loss rate. 

\begin{figure}
\vspace{0cm}
\epsfxsize=8.5cm 
\epsfbox{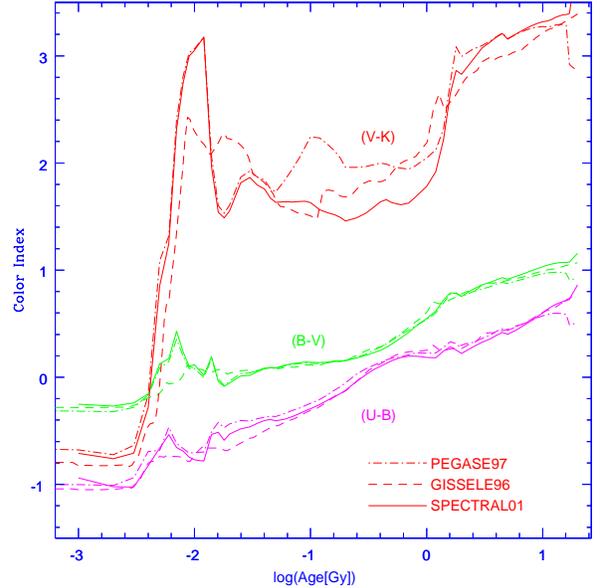} 
\vspace{0cm}
		 {\caption[]{Predicted color index evolution of star-burst at 
$Z_\odot$ by three different codes; PEGASE97, GISSEL96, and {\code}.}
\label{f3}}
\end{figure}

1) In the first model we compare the evolution of colors at $Z=Z_\odot$ with 
those obtained from PEGASE97 (\cite{pegase}, FR) and GISSEL96 (\cite{bch}, 
BC). The Figure \ref{f3} shows this comparison.

We can see from the Figure \ref{f3}, that there is a good agreement between 
the three models for colors. The differences are due to the use of different 
spectral library (PEGASE97), or different stellar 
tracks (GISSEL96). In the same Figure \ref{f3} it is possible to see a pick 
in the $(V - K)$ color, around $log(Age[Gy])=-1.0$, which is produced by 
thermal pulses generated by stars in range ($5.0 \leq m_i/M_\odot \leq 
10.0$), explained by FR. For a young population (around $log(Age[Gy])=-2.0$) 
we can see an agreement between our model and that from FR in colors, and the 
agreement occurs with colors from BC in the inverse case when we compare old 
populations (around $log(Age[Gy])=1.0$). 

\begin{figure}
\vspace{0cm}
\epsfxsize=8.5cm 
\epsfbox{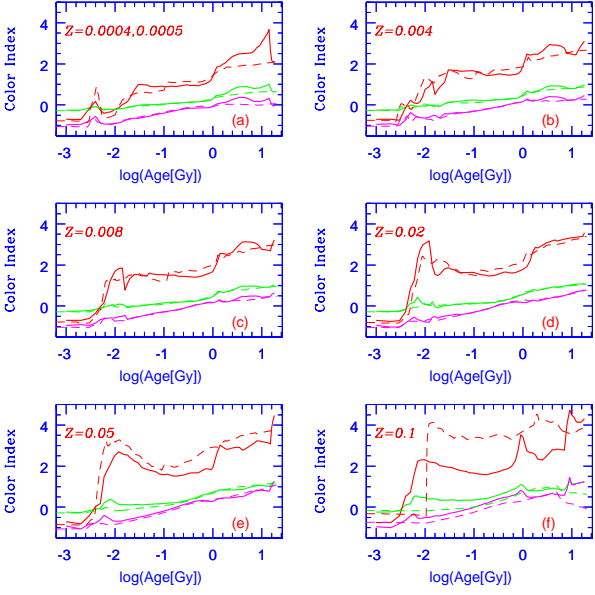} 
\vspace{0cm}
		 {\caption[]{Colors evolution as Figure \ref{f3} 
for star-bursts at different metallicities obtained with GISSEL96 (dashed 
lines) and {\code} (full lines).}
\label{f4}}
\end{figure}

\begin{figure}
\vspace{0cm}
\epsfxsize=8.5cm 
\epsfbox{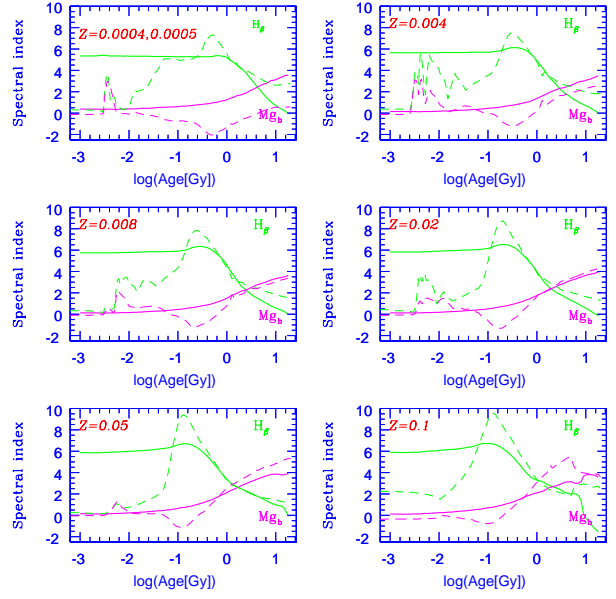} 
\vspace{0cm}
		 {\caption[]{Spectral indices evolution  as the Figure 
\ref{f4} for star-bursts at different metallicities. GISSEL96 (dashed 
lines), and {\code} (full lines).}
\label{f5}}
\end{figure}

2) In the second model we compare the same colors showed in Figure \ref{f3} 
for different metallicities with those obtained from GISSEL96 
(Figure \ref{f4}). In the same way we compare the evolution of two Lick 
indices $H_\beta$ y $Mg_b$ in Figure \ref{f5}.

In the plot (a) of the Figure \ref{f4} we have compared models for the lowest 
metallicity reached linearly by our code with models at lowest metallicity in 
Padova tracks used by GISSEL96. There is a good agreement between models 
with different metallicities except for those at $Z=0.1$ in plot (f) because 
the original tracks are incomplete. 

In the case of indices, our models predict a flat and constant evolution for 
young populations (around $log(Age[Gy]) \leq -1.0$), while indices from BC 
present variations. The differences are due to they extend linearly to high 
temperatures the fitting functions (depending on $T_{eff}$, $log(M/H)$, 
and $log(g)$), which predict the value of each index, and we use the original 
fitting functions with temperature limit around 13,000 K. There is an 
agreement between both models for populations in the range of 
$log(Age[Gy]) > -1.0$.

\subsubsection{Spectra}

We have tested spectra in a different way. We have divided the test for two 
ranges, for young stellar populations and for old stellar populations. Both 
ranges were subdivided in three metallicities. We use instantaneous 
star-bursts as the star formation rate.

For {\bf young populations} we assume the next set of conditions:

An $\Phi(m_i) \propto m^{\alpha}$, with $\alpha=-2.35$, low mass limit, 
$m_l=1.0$, and high mass limit, $m_h=100.0$, and for {\code} we use Geneva 
tracks with high mass loss rate. 

\begin{figure}
\vspace{0cm}
\epsfxsize=8.5cm 
\epsfbox{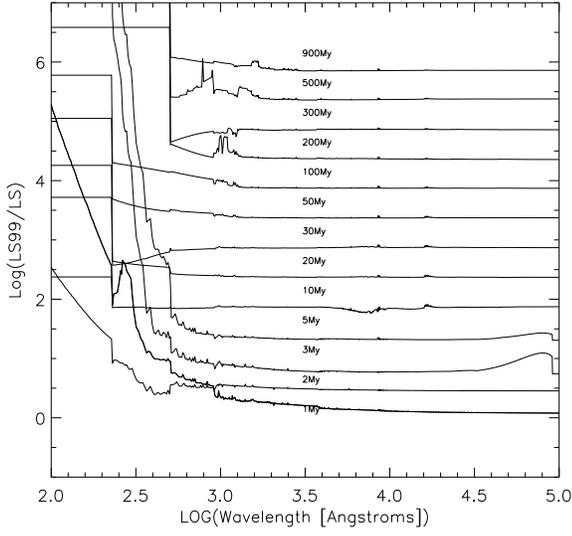} 
\vspace{0cm}
		 {\caption[]{Quantitative differences between our spectra and 
those from STARBURST99 at different ages, for a star-burst at $Z=Z_\odot$.}
\label{f6}}
\end{figure}

1) In the first model we compare our spectra with different ages in the range 
of $10^{-3} \leq t/Gy \leq 0.9$ at $Z=Z_\odot$ with spectra produced by 
STARBURST99 (\cite{leitherer}). Figure \ref{f6} shows quantitative 
differences taking the $log(L_{{S99}_k}/L_{S_k}) + 0.5k$ 
($k=1,13$), where $L_{S99}$ is the 
luminosity from STARBURST99's spectra and $L_S$ from {\code}. All models from 
STARBURST99 used here are without nebular emission.

There is a very good agreement, and it is comprehensive, because both of 
codes use the same spectra library and the same stellar tracks. There is a 
contribution in the UV part of spectra for ages at 3 and 5 My because the 
WR phases at that ages, that is the reason why the luminosity increase in UV 
part and the rate of luminosities between both sets of models is higher. The 
same features are not visualized in our spectra because the library for 
massive stars with strong stellar winds are not considered in our spectra 
library. This increase of luminosity in UV part appears for STARBURST99 
models at high metallicity, but not for low metallicity and the rate of 
luminosities is lower for low $Z$.

\begin{figure}
\vspace{0cm}
\epsfxsize=8.5cm 
\epsfbox{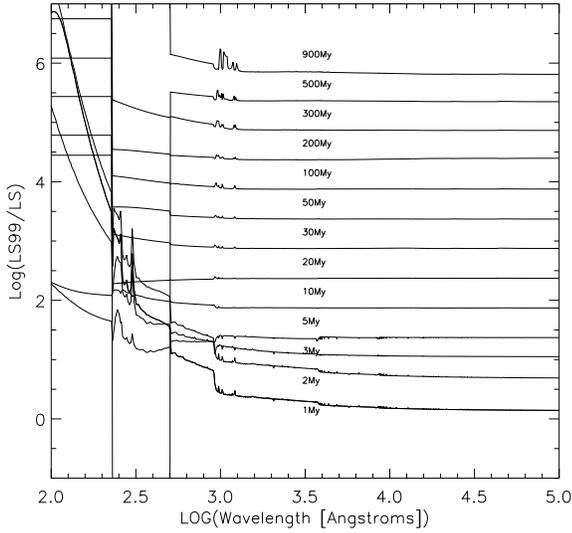} 
\vspace{0cm}
		 {\caption[]{Quantitative differences between our spectra and 
those from STARBURST99 at different ages, for a star-burst at $Z=0.001$.}
\label{f7}}
\end{figure}

2) In the same way, the second model compares spectra with different ages at 
$Z=0.001$. Figure \ref{f7} shows quantitative differences of these models.

\begin{figure}
\vspace{0cm}
\epsfxsize=8.5cm 
\epsfbox{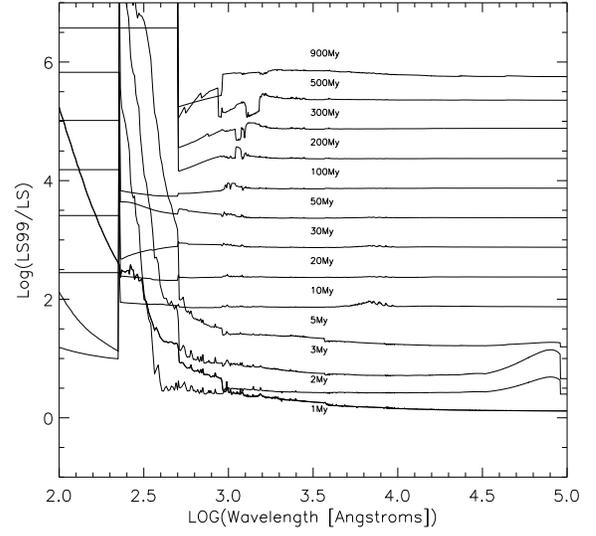} 
\vspace{0cm}
		 {\caption[]{Quantitative differences between our spectra and 
those from STARBURST99 at different ages, for a star-burst at $Z=0.04$.}
\label{f8}}
\end{figure}

3) The final comparison is done for spectra at $Z=0.04$. Figure \ref{f8} 
shows quantitative differences.

These three sets of models show a very good agreement with models from 
Leitherer et al. (1999). The same features for the case at $Z_\odot$ remain in 
models (2) and (3). This agreement shown in three different 
star-burst populations at different metallicities is present in the rest of 
metallicities of Geneva tracks as showed in \cite{vazquezth}.

For {\bf old populations} we assume the next set of conditions:

An $\Phi(m_i) \propto m^{\alpha}$, with $\alpha=-2.35$, low mass limit, 
$m_l=0.1$, and high mass limit, $m_h=100.0$; and for {\code} we use Geneva 
tracks with normal mass loss rate. 

\begin{figure}
\vspace{0cm}
\epsfxsize=8.5cm 
\epsfbox{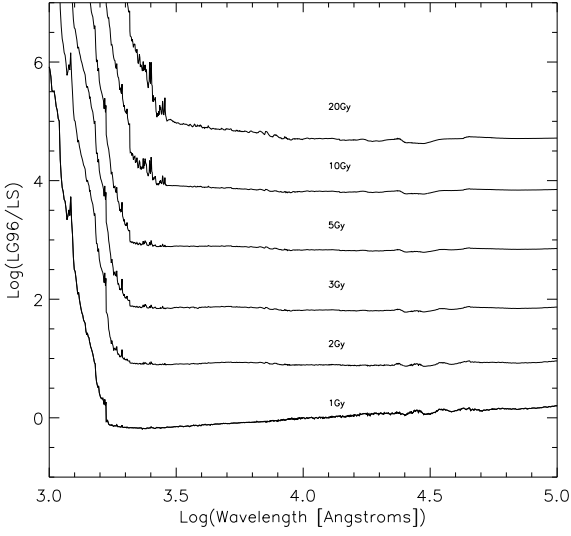} 
\vspace{0cm}
		 {\caption[]{Quantitative differences between our spectra and 
those from GISSEL96 at different ages, for a star-burst at $Z=Z_\odot$.}
\label{f9}}
\end{figure}

1) In the first model we compare our spectra with different ages in the range 
of $1.0 \leq t/Gy \leq 20.0$ at $Z=Z_\odot$ with spectra produced by 
GISSEL96 (BC). Figure \ref{f9} shows quantitative differences taking the 
$log(L_{G96_k}/L_{S_k}) + k$ ($k=1,6$), where $L_{G96}$ is 
the luminosity from GISSEL96's spectra and $L_S$ from {\code}.

It is easy to see some differences from Figure \ref{f9}, which are due to 
the stellar tracks used for each code, Padova (GISSEL96) and Geneva 
({\code}), respectively. The considered spectral library is the same, 
but not the same version, which is a source of differences as well. The main 
differences are in the blue part of the spectra, where we find a contribution 
in UV due to old stars in phases like HB and planetary nebula predicted by 
additional stellar library used by GISSEL96, and there is not a counterpart 
in our models. Again, this is the reason why the rate of luminosities is 
higher for those models from GISSEL96.

\begin{figure}
\vspace{0cm}
\epsfxsize=8.5cm 
\epsfbox{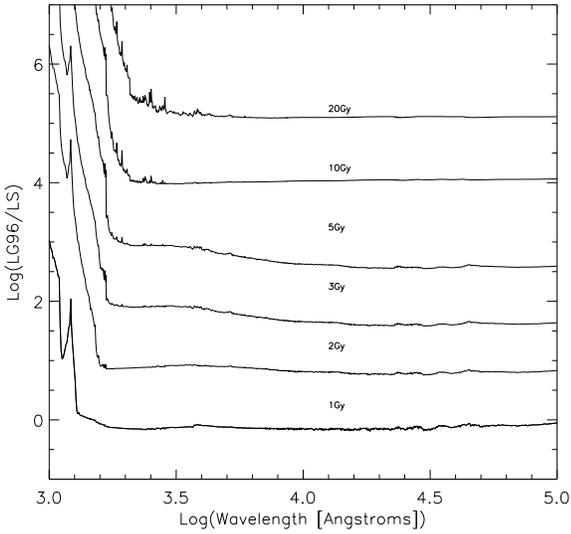} 
\vspace{0cm}
		 {\caption[]{Quantitative differences between our spectra and 
those from GISSEL96 at different ages, for a star-burst at $Z=0.004$.}
\label{f10}}
\end{figure}

2) In the same way, the second model compares spectra with different ages at 
$Z=0.004$. Figure \ref{f10} shows quantitative differences of these models.

\begin{figure}
\vspace{0cm}
\epsfxsize=8.5cm 
\epsfbox{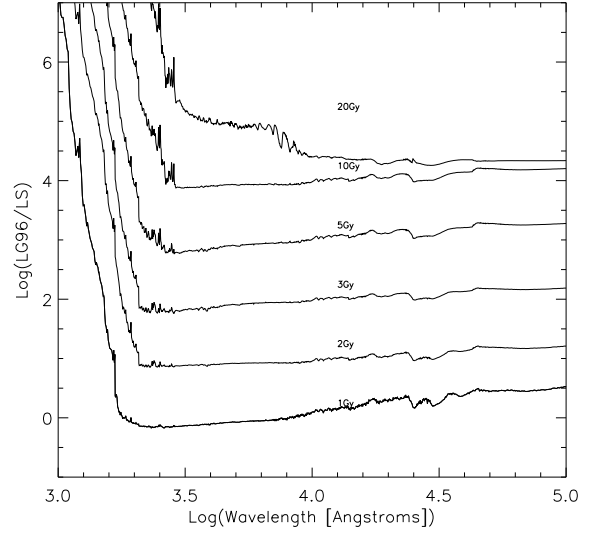} 
\vspace{0cm}
		 {\caption[]{Quantitative differences between our spectra and 
those from GISSEL96 at different ages, for a star-burst at $Z=0.05$.}
\label{f11}}
\end{figure}

3) The final comparison is done for spectra at $Z=0.05$. Figure \ref{f11} 
shows quantitative differences.

The features shown in models (2) and (3) are explained in the same form as 
those with $Z_\odot$. The rest of models at different metallicities are shown 
in \cite{vazquezth}.

\subsection{Models for the irregular galaxy NGC 1560}

Carigi, Col\'{\i}n \& Peimbert (1999) discussed the chemical evolution of
three irregular galaxies: NGC 1560, II Zw 33 and a mean irregular galaxy. In
the attempt to apply spectro-chemical evolution models to this three galaxies,
we look for spectro-photometric measurements of these galaxies in the 
literature. In this work, we compute spectro-chemical evolution models only 
to NGC 1560 because: i) For NGC 1560 and II Zw 33 the nonbaryonic dark matter
inside the Holmerg Radius is known, which constrains the chemical evolution
models. ii) For NGC 1560, but nor for II Zw 33, neither the sample galaxies 
that form the mean irregular galaxy, there are photometric measurements to 
constrain the spectral models.

We have used close box models from \cite{carigi99} updated with yields from 
\cite{maeder} and \cite{vanden} to be consistent with stellar evolution 
models with normal mass loss rate used by our code. Chemical evolution models 
reproduce two constraints O/H abundance (\cite{richer}) and 
$\mu=M_{gas}/M_{total}$ (\cite{walter}, \cite{broelis}). Three models with 
different ages and star formation rates to match the gas metallicity observed 
are not able to give us information about the age of star formation history. 
Colors of this galaxy may constrain the age if we use evolutionary synthesis.

These three chemical models results were used by our code to constrain the 
stellar population age which dominates the light in the irregular galaxy NGC 
1560.

These models use the basic form for the IMF from Kroupa, Tout \& Gilmore 
(1993), in the range of $0.01 \leq m_i/M_\odot \leq 120.0$ as follow:

\begin{equation}
\Phi(m_i)  \sim \left\{
\begin{array}{l}
\mbox{$m_i^{- \alpha}$\ \ \ \ \ \ $if \ \ 0.01 \leq m_i/M_\odot < 0.5,$}\\
\mbox{$m_i^{-2.2}$\ \ \ \ $if \ \ 0.5 \leq m_i/M_\odot < 1.0,$}\\
\mbox{$m_i^{-2.7}$\ \ \ \ $if \ \ 1.0 \leq m_i/M_\odot < 120.0$}\\
\end{array}
\right.
\end{equation}
where $\alpha$ depends on chemical constraints and age of the model, and had 
three different values for every model:

\begin{equation}
\alpha = \left\{
\begin{array}{l}
\mbox{$-2.21$\ \ \ \ 0.1 Gy Model}\\
\mbox{$-2.24$\ \ \ \ 1.0 Gy Model}\\
\mbox{$-2.27$\ \ \ \ 10.0 Gy Model}\\
\end{array}
\right.
\end{equation}

\begin{figure}
\vspace{0cm}
\epsfxsize=8.5cm 
\epsfbox{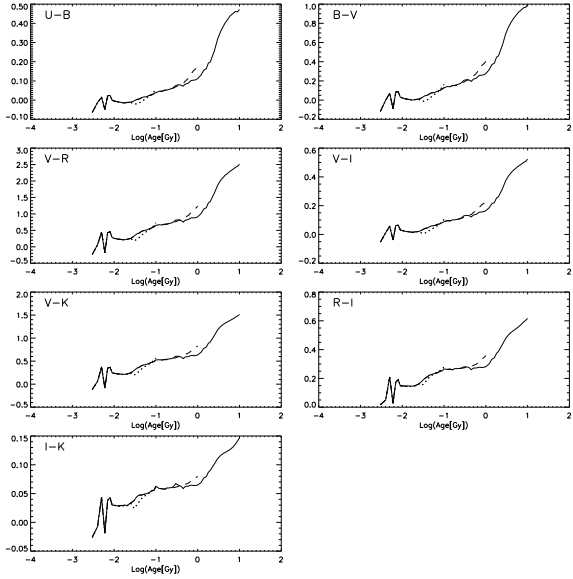} 
\vspace{0cm}
		 {\caption[]{Broad band colors evolution for the three models 
at 0.1 Gy (dotted line), 1.0 Gy (dashed line), and 10.0 Gy (continuous line).}
\label{f12}}
\end{figure}

We use $\Phi(m_i)$, $\Psi(t)$, and $Z(t)$ from chemical models in {\code} to 
produce a set of models. The set was produced by using tracks with the normal 
mass loss rate. Figure \ref{f12} shows the evolution of some broad band 
colors predicted for this galaxy.

\begin{figure}
\vspace{0cm}
\epsfxsize=8.5cm 
\epsfbox{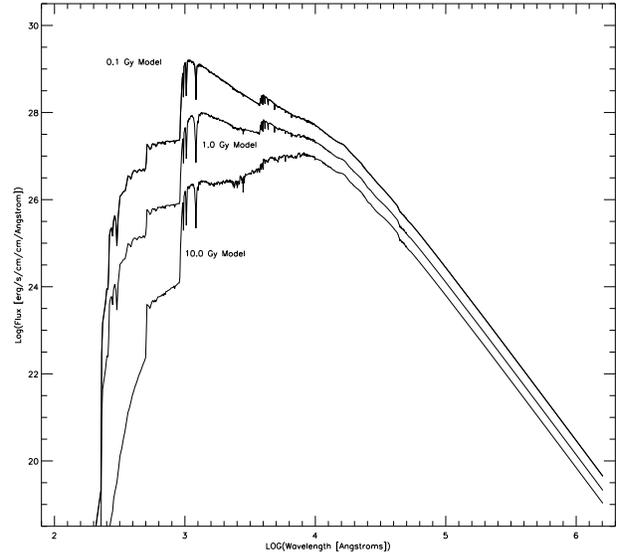} 
\vspace{0cm}
		 {\caption[]{The spectra produced for the three models.}
\label{f13}}
\end{figure}

The rest of broad band colors and Lick index for this set of models are shown 
in \cite{vazquezth}. The mean spectra produced for the models are shown in 
Figure \ref{f13}.

\begin{figure}
\vspace{0cm}
\epsfxsize=8.5cm 
\epsfbox{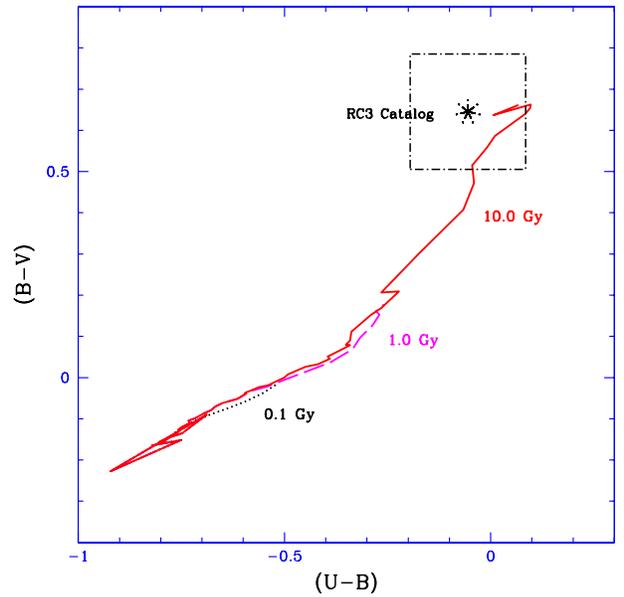} 
\vspace{0cm}
		 {\caption[]{($U - B$) vs. ($B - V$) evolution for the three 
models are compared with observed values. Dotted line shows the 0.1 Gy model, 
dashed line for the 1.0 Gy model, and continuous line for the 10.0 Gy model. 
Box shows the error for the values from the literature.}
\label{f14}}
\end{figure}

The main goal in this work is to compare our predictions for this galaxy with 
those obtained observationally. For this galaxy we have obtained $(U - B)_0$ 
and $(B - V)_0$ from the \cite{rc3}, which are mean 
values. The evolution and the comparison is shown in Figure \ref{f14}.

\begin{figure}
\vspace{0cm}
\epsfxsize=8.5cm 
\epsfbox{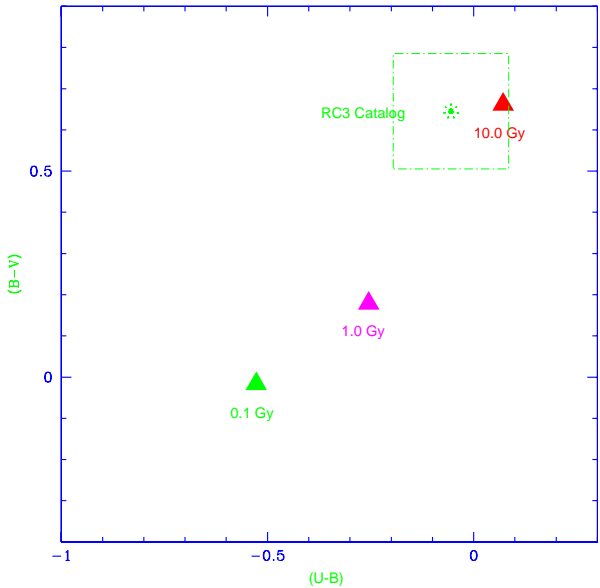} 
\vspace{0cm}
		 {\caption[]{($U - B$) vs. ($B - V$) from models 
compared with values from the literature.}
\label{f15}}
\end{figure}

Figure \ref{f15} shows that the model which has a star 
formation age of 10.0 Gy fits in the best way the observations, it does not 
occur with younger models. 

A better fit would be obtained if there were more colors, indices 
or the full spectrum in the literature.

\section{Conclusions}

We have extended the tracks from the Geneva group, producing two sets of 
tracks, one with high mass loss rate for masses in the range of 
$12.0 \leq m_i/M_\odot \leq 120.0$ and the other with normal mass loss rate 
in the range of $0.08 \leq m_i/M_\odot \leq 120.0$. These sets of tracks are 
complete and consistent for metallicities in the range 
$0.001 \leq Z \leq 0.04$, they cover phases from main sequence up to 
asymptotic giant branch in the range $0.8 \leq m_i/M_\odot \leq 120.0$, and 
pre-main sequence and main sequence for masses in 
$0.08 \leq m_i/M_\odot \leq 0.7$. Together with the spectral library, the 
code {\code} is able to predict the spectral properties from stellar 
populations under a wide variety of conditions.

With this code, it is possible to build synthetic populations as the 
sophisticated chemical evolution models predict, as long as metallicity is in 
the range $0.001 \leq Z \leq 0.04$.
 
Results for star-burst evolution in broad band colors, spectral indices and 
spectra from our code fit very well with the same results obtained with other 
codes. {\code} could be used in young stellar populations like in star-burst 
or blue galaxies, or stellar populations inside cores of AGN's. In the same 
way, the code could be used for old stellar populations like those in early 
type galaxies and all galaxy types in the middle of young and old stellar 
populations.

In applying just chemical models using observational constraints mentioned 
here for NGC 1560 it is not easy to distinguish what it is the age of the 
population. But using observed colors for this galaxy a spectro-chemical 
evolution model can be fitted and a population around 10 Gy and mean 
metallicity of stars $Z=0.002$ are predicted with our model.

In the same way, from Figure \ref{f14}, the evolution of colors for the 
10.0 Gy model is possible to see that colors match the box error at 5 Gy. 
Furthermore, we might say that star formation history age for the mean stellar 
population is between [5.0,10.0 Gy], though we do not have chemical model for 
5 Gy. 

We have obtained success in reproducing the values for many simple star 
formation scenarios, and finally, we have matched very closely the values 
observed for two broad band colors in NGC 1560.

\begin{acknowledgements}

This work is part of the PhD thesis developed by Gerardo A. V\'azquez in the 
IA-UNAM who thanks to Gustavo Bruzual made an extensive review to the code 
developed in his thesis. Gerardo is similarly indebted to Gloria 
Koenigsberger for the final push to finish his PhD. We thank the Instituto de 
Astronom\'{\i}a UNAM for financial support of Gerardo with a fellowship, 
resources and travels obtained through the projects PAPIIT, IN126098 and 
IN109696. We thanks to Gloria Koenigsberger, Luc Binette and Claus Leitherer 
for reading the manuscript.

\end{acknowledgements}

\end{document}